\documentclass{llncs}

\usepackage{amsmath}
\usepackage{amssymb}
\usepackage{graphicx}
\usepackage{caption}
\usepackage{verbatim}

\title{S4: A New Secure Scheme for Enforcing Privacy\\ in Cloud Data Warehouses}

\author{Somayeh Sobati Moghadam%\inst{1} 
\and J\'{e}r\^{o}me Darmont%\inst{1} 
\and G\'{e}rald Gavin%\inst{2}
}

\institute{
Universit\'{e} de Lyon, Lyon 2, Lyon 1, ERIC EA3083\\
5 avenue Pierre Mend\`{e}s France -- 69676 Bron Cedex -- France\\
\email{ssobati@eric.univ-lyon2.fr, jerome.darmont@univ-lyon2.fr, gerald.gavin@univ-lyon1.fr}
}

\begin{document}

\maketitle              % typeset the title of the contribution

\begin{abstract}
Outsourcing data into the cloud becomes popular thanks to the pay-as-you-go paradigm. However, such practice raises privacy concerns. The conventional way to achieve data privacy is to encrypt  sensitive data before outsourcing. When data are encrypted, a trade-off must be achieved between security and efficient query processing. Existing solutions that adopt multiple encryption schemes induce a heavy overhead in terms of data storage and query performance, and are not suited for cloud data warehouses.  
In this paper, we propose an efficient additive encryption scheme (S4) based on Shamir's secret sharing for securing data warehouses in the cloud. S4 addresses the shortcomings of existing approaches by reducing overhead while still enforcing good data privacy. Experimental results show the efficiency of S4 in terms of computation and storage overhead with respect to existing solutions. 
%\keywords{Cloud computing, Data Warehouses, Data Privacy, Encryption Schemes, Aggregation Queries}
\end{abstract}

\section{Introduction}

Data warehouses (DWs) provide a consolidated view of organizations and businesses' data,
optimized for reporting and analysis. %Hence, DWs can
greatly enhance decision making. 
DWs consolidate historical data from different sources and allow on-line  analytical processing (OLAP).
Nowadays, data outsourcing scenarios tremendously grow with the advent of cloud computing that offers both %. Cloud computing is appealing because of 
cost savings and service benefits. %Moreover, cloud computing provides higher availability, scalability and more effective disaster recovery than in-house operations \cite{DBLP:conf/storagess/DamianiVFJPS05}. 
One of the most notable cloud outsourcing services is Database-as-a-Service, where individuals and organizations outsource data storage and management to a Cloud Service Provider (CSP) \cite{DBLP:journals/toit/XiongCL07}. Naturally, such services allow outsourcing a DW and running OLAP queries \cite{DBLP:conf/dbsec/AmanatidisBO07}.
%Although 
Yet, data outsourcing %induces many benefits, it also 
brings out privacy concerns %. Privacy issues arise 
since sensitive data are stored, maintained and processed by an external third party that may not be fully trusted.

A typical solution to preserve data privacy is encrypting data locally before sending them to an external server. Secure database management systems (SDBMSs) such as CryptDB \cite{DBLP:conf/sosp/PopaRZB11} implement cryptographic schemes. % such as order-preserving or homomorphic encryption, and allow querying encrypted data without decryption. 
Paillier's partially homomorphic encryption scheme \cite{paillier1999public} is notably used in CryptDB to provide high security. However, it induces a high storage and computation overhead. %that negatively affects query processing. 
Hence, in this paper, we propose a new Secure Secret Splitting Scheme (S4) that aims at replacing Paillier's scheme in systems such as CryptDB. S4 is based on the idea of secret sharing \cite{shamir1979share} %. S4 implements an additive homomorphic scheme, i.e., additions can be directly computed over encrypted data. S4 
and is efficient both in terms of storage and computing, without sacrificing privacy too much. %, which is ideal for data outsourcing scenarios that consider the user has limited computation and storage resources.

In the remainder of 
this paper, % is organized as follows. 
Section~\ref{Section.RelatedWork} discusses related works about SDBMSs, homomorphic encryption and secret sharing. %Section \ref{Section.Background} introduces secret sharing, a particular encryption technique that strongly inspires S4. 
Section \ref{Section.S4} details and discusses S4. Section \ref{Section.Experimental Evaluation} provides an experimental validation of S4 against Paillier's scheme. Finally, section \ref{Section.Conclusion} concludes the paper and hints as future research.

\section{Related Works}
\label{Section.RelatedWork}

\subsection{Secure Database Management Systems}

%\subsubsection{CryptDB}
%\label{section.RelatedWorks:CryptDB}

CryptDB brings together powerful cryptographic tools to handle query processing on encrypted data without decryption \cite{DBLP:conf/sosp/PopaRZB11}. 
Encryption in CryptDB is like onion layers that store multiple ciphertexts, i.e., encrypted data, within each other. 
Each onion layer enables certain kind of query processing and a given security level provided by one encryption scheme. For instance, order-preserving encryption (OPE) enables range queries and additive homomorphic encryption %(Section \ref{Section:RelatedWorks:HE}) 
enables addition over encrypted data.
%Encryption in CryptDB is based on the following encryption schemes. Random (RND) bears the highest security as it provides semantic security (indistinguishably under an adaptive chosen plaintext, i.e., clear data, attack). Deterministic (DET) allows equality checking. Order-preserving encryption (OPE) enables range queries. Additive homomorphic encryption (HOM) (section \ref{Section:RelatedWorks:HE}) enables addition over encrypted data, and keyword search (SEARCH) over encrypted texts. The outermost layers RND and HOM provide the highest level of security, whereas inner layers, i.e., OPE and DET, provide more functionality. 
Yet, CryptDB is not perfectly secure since schemes such as OPE
%and introduces some loopholes.  Database attributes that support either range or exact match queries are encrypted
%with OPE or DET, respectively. They are thus vulnerable to statistical attacks, because OPE and DET 
reveal some statistical information about plaintext \cite{DBLP:conf/ccs/NaveedKW15}. %Moreover, while CryptDB supports many SQL queries, there are still a lot of queries that are not supported, e.g., predicate evaluation on more than one attribute is not possible. As a result, CryptDB supports only 2 queries out of 22 from the TPC-H decision-support benchmark \cite{TPCH}.

%\subsubsection{MONOMI}
%\label{section.RelatedWorks:MONOMI}

MONOMI builds upon CryptDB to allow the execution of analytical workloads over encrypted data outsourced to the cloud \cite{DBLP:journals/pvldb/TuKMZ13}. MONOMI aims at improving CryptDB's query processing capability and efficiency based on split client/server execution. %It introduces a query  planner that splits query execution between the server and client. 
%Moreover, 
A designer also optimizes physical data layout. % at the server's. Some optimization techniques such as per-row computation, space-efficient encryption, grouped homomorphic addition, and pre-filtering improve the performance of query processing and allow more analytical queries to be processed.  

%MONOMI allows more query types than CryptDB, e.g., 19 TPC-H queries out of 22, but induces a heavy communication load between user and server. For instance, intermediate results may be exchanged several times to execute different parts of a query  \cite{DBLP:journals/pvldb/TuKMZ13}. 

%\subsubsection{Trusted Hardware}

Eventually, using a local trusted hardware at the CSP's, such as TrustedDB \cite{DBLP:conf/sigmod/BajajS11} and CipherBase \cite{DBLP:conf/icde/ArasuEJKKR15},
is an alternative approach to query encrypted data.
%The idea is processing queries inside tamper-proof enclosures of trusted hardware. These components have access to database keys, allow performing a limited set of queries over encrypted data and are physically hosted at the CSP's.
However, trusted hardware is limited in computation ability and memory capacity, and also very expensive. %, which is contrary to using cheap commodity machines in the cloud.

\subsection{Homomorphic Encryption }
\label{Section:RelatedWorks:HE}

%\subsubsection{Fully Homomorphic Encryption}

Fully homomorphic encryption (FHE) allows performing arbitrary arithmetic operations over encrypted data without decryption \cite{gentry2009fully}. FHE provides semantic security, i.e., it is computationally impossible to distinguish two ciphertexts encrypted from the same plaintext.
However, FHE requires so much computing power that it cannot be used in practice. 

%\subsubsection{Partially Homomorphic Encryption}

Partially homomorphic encryption (PHE) is more efficient than FHE. % and closer to practice implementation. It allows either addition or multiplication over encrypted data, but not both. %If a PHE scheme allows addition or multiplication, it is referred to as additive homomorphic and multiplicative homomorphic, respectively. PHE  
Paillier's \cite{paillier1999public} %is an example of  additive homomorphic scheme and is currently 
the most efficient additive FHE. 
With Paillier's scheme, multiplying the encryption of two values results in an encryption of the sum of the values, i.e., 
$Enc_k(x) \times Enc_k(y) = Enc_k(x+y)$, where the multiplication is performed modulo some public-key $k$ \cite{DBLP:conf/sosp/PopaRZB11}. %Paillier's cryptosystem is used in CryptDB and MONOMI to compute \texttt{SUM} and \texttt{AVG} aggregation queries over encrypted data without decryption. %CryptDB replaces SUM with calls to user defined functions (UDF) at the SP's that performs Paillier's multiplication on a column encrypted with PHE. \par		
Paillier's scheme is, however, still computationally intensive and induces as large ciphertext sizes as %. It indeed operates over large plaintext and ciphertext values such as 1024 and
2048 bits. Additionally, modular multiplications become computationally expensive on a %, especially in the case of a 
large number of records, such as in the fact table of a DW \cite{DBLP:journals/pvldb/TuKMZ13,DBLP:reference/dbsec/Sion08}. 

%Some solutions have been proposed to improve the efficiency of Paillier's scheme. Packing multiple integer values into a single Paillier plaintext and grouped homomorphic addition techniques \cite{ge2007answering} are used in MONOMI. To make efficient use of plaintext payload, MONOMI packs values from multiple rows into a single Paillier plaintext \cite{DBLP:journals/pvldb/TuKMZ13}. However, such packing techniques make impossible partial operations or query processing over values, e.g., when only some values are needed. 

%Grouped homomorphic addition packs all attributes that are aggregated together (e.g., TPC-H query 1 requires multiple aggregations over multiple attributes) in such a way that aggregation queries can be computed with a single modular multiplication. However, grouped homomorphic addition requires that all queries are declared by the application ahead of time, which it is not possible for all applications. In essence, MONOMI aims at TPC-H's queries and cannot be applied for real applications, especially for OLAP ad-hoc navigation. 

%\section{Background}
%\label{Section.Background}

%\subsection{Shamir's Secret Sharing}
\subsection{Secret Sharing}

Secret sharing divides a secret piece of data into so-called shares that are stored at $n$ participants'. %One single participant has no means to reconstruct the secret; 
A subset of $k\leq n$ participants is required to reconstruct the secret.
%Shamir's, the first secret sharing scheme, % is based on polynomial interpolation . It 
%works by defining a polynomial of degree $k-1$ with the secret as the constant term \cite{shamir1979share}. Then, $n\geq k$ points are built on the polynomial. Since at least $k$ points are needed to uniquely reconstruct a polynomial of degree $k-1$, knowing $k$ points, the secret can be reconstructed using Lagrange interpolation \cite{DBLP:conf/dbsec/DautrichR12}. %In Shamir's scheme, the  points and coefficients are taken from a finite field $\mathbb{F}_p$, where $p$ is a prime.  
In Shamir's, the first secret sharing scheme \cite{shamir1979share},
to share a secret $v_j$, a random polynomial $P_{v_j}(x)$ of degree $k-1$ is first built. 
The owner of the secret chooses a prime $p>v_j$ and $k-1$ random numbers $a_1, a_2,..., a_{k-1}$ from $\mathbb{F}_p$; and sets $a_0=v_j$ (Equation \ref{Equation.Poly}). $P_{v_j}(x)$ passes through the point $(0,v_j)$. 

\begin{equation}\label{Equation.Poly}
P_{v_j}(x) = a_{k-1}x^{k-1} + ...+ a_1x+a_0 \quad \mod p
\end{equation}

To build $n$ points over $P_{v_j}(x)$, a set of $n$ distinct elements in $\mathbb{F}_p$, $X=\{ \mathtt{x_1},  \mathtt{x_2},\dots,  \mathtt{x_n}\}$, is chosen such that $ \mathtt{x_i} \neq0$ $\forall i=1,...,n.$
For each participant $i$, the corresponding share is $v_{i,j}=P_{v_j}( \mathtt{x_i})$. For each secret $v_j$, there are $n$ points $(\mathtt{x_i}, v_{i,j})$ through which the polynomial $P_{v_j}(x)$ passes \cite{DBLP:journals/ijisec/HadaviJDC15}. 
%Note that $X$ must be highly protected from untrusted parties.
Any $k$ shares form $k$ points $(\mathtt{x_i}, v_{i,j})$ $i=1,\dots k$, from which polynomial $P_{v_j}(x)$ can be reconstructed  using Lagrange interpolation \cite{DBLP:conf/dbsec/DautrichR12} (Equation \ref{Equation.Lagrange11}). 

\begin{equation}\label{Equation.Lagrange11}
P_{v_j}(x) = \sum_{i=1}^{k} v_{i,j}\ell_i(x) \quad \mod p
\end{equation}

%where $\ell_i(x)$ is the Lagrange basis polynomial:

\begin{equation*}\label{Equation.Lagrange2}
\ell_i(x)=\prod_{ 1 \leq j \leq k,j\neq i}^{}(x - \mathtt{x_j})(\mathtt{x_i} - \mathtt{x_j})^{-1} \quad \mod p  
\end{equation*} 

where $(\mathtt{x_i} - \mathtt{x_j})^{-1}$ is the multiplicative inverse of $(\mathtt{x_i} - \mathtt{x_j})$ modulo $p$ \cite{DBLP:conf/dbsec/DautrichR12}.
Eventually, the secret is the constant term of the polynomial:

\begin{equation}\label{Equation:LagrangeBasis}
v_j=P_{v_j}(0)=\sum_{i=1}^{k} v_{i,j}\ell_i(0) \quad \mod p.
\end{equation}

\section{S4}
\label{Section.S4}

\begin{center}
	\begin{figure}[hbt]
		%%\vspace{2.85cm}
		\centering
		\includegraphics[width=7cm]{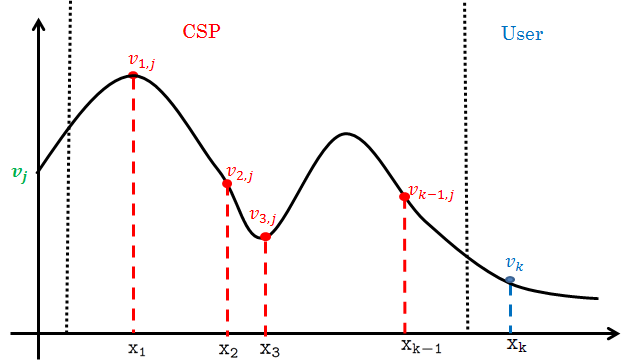}
		\caption{S4 secret splitting}
		\label{fig:Poly}
	\end{figure}
\end{center}

%S4 is a new scheme to securely store data, while allowing  summation queries to be computed at a CSP's efficiently. 
S4's driving idea is based on secret sharing, but instead of sharing secrets to $n$ participants' or CSPs', they are stored at one single CSP's.
%. Yet, in contrast to similar solutions that used secret sharing to outsource data, shares are stored at only one CSP's. 
Thus, we avoid the high storage overhead of secret sharing. %induced by sharing data at $n$ participants'/CSPs'.
In S4, each secret $v_j$ is divided into $n=k$ splits $v_{1,j},...,v_{k,j}$. %Since shares are not actually distributed to several participants, we call them splits. 
$k-1$ splits,  $v_{1,j},...,v_{k-1,j}$, are stored at the CSP's and $v_{k,j}$ is stored in a trusted machine, e.g., at the user's (Figure~\ref{fig:Poly}). In order to reduce storage overhead at the user's, $v_{k,j}$ is set to be the same for all secrets. 

\subsection{Splitting and Reconstruction Processes}\label{Section.S4.SplittingProcess}

First, 
$\mathtt{x_k}$ and $v_k$ are randomly set up from  $\mathbb{F}_p$, where $p$ is a big prime number, i.e., greater than the greatest possible query answer.  
For any secret $v_j$, a random polynomial $P_{v_j}(x)$ is built that passes through $(0,v_j)$ and $(\mathtt{x_k}, v_k)$.
To this end, $k-2$ points $(a_i, b_i), i=1,..., k-2$ are chosen randomly from $\mathbb{F}_p$ 
such that $a_i \neq \mathtt{x_k}$ and $a_i \neq 0$  $\forall i=1,...,k-2$.
Given $k$ points $(a_1, b_1), (a_2, b_2), ... , (a_{k-2}, b_{k-2})$, $(0, v_j)$ and $(\mathtt{x_k}, v_k)$, polynomial $P_{v_j}(x)$ is built using Equation~\ref{Equation.Lagrange11}. %Note that it is not necessary to 
Storing the $k-2$ random points is unnecessary because they are not needed for secret reconstruction.  

To divide $v_j$ into $k-1$ splits (since $(\mathtt{x_k},v_k)$ is already fixed), a  set of $k-1$ distinct elements  $X=\{\mathtt{x_1}, \mathtt{x_2}, \dots, \mathtt{x_{k-1}}\}$ is chosen from $\mathbb{F}_p$ such that $\mathtt{x_i} \neq 0$ and $\mathtt{x_i} \neq \mathtt{x_k}$  $\forall i=1, ..., k-1$. 
Then, splits are $v_{i,j}=P_{v_j}(\mathtt{x_i})$. 
$\mathcal{K}$=$(X, (\mathtt{x_k},v_k))$ is considered as a \textit{private key} for S4 and must be %highly protected and 
kept hidden from the CSP.
%As in secret sharing, the key observation is that $k$ points  are required to build a unique polynomial of degree $k-1$. 
To reconstruct secret $v_j$, its $k-1$ splits must be retrieved from the CSP. Given points $(\mathtt{x_i}, v_{i,j})$, $i=1,...,k-1$ and  $(\mathtt{x_k}, v_k)$, which is stored at the user's, polynomial $P_{v_j}(x)$ can be reconstructed using Equation~\ref{Equation.Lagrange11}. Its constant term is $v_j$.

\subsection{Summation Queries}
\label{Section.S4.AggregationQueries}

Let a relational table $T$ consist of one attribute $A$ (additional attributes, if any, can be processed similarly). Suppose $T$ has $m$ records. We denote by $v_j$ the $j^{th}$ value of $A$. 
For attribute $A$ in $T$, $k-1$ attributes $A_i$, $i=1,..., k-1$ are created in table $T^\prime$ at the CSP's, where each attribute $A_i$ stores the $i^{th}$ splits. 
Without loss of generality, we assume integer data type for $A$. Other data types %, e.g., reals, characters, strings, 
can be transformed into integers before splitting.
S4 allows summation queries to be computed directly at the CSP's.   
Consider a query that sums $q$ values of $A$.

\begin{equation*}\label{Equation.sumi}
\texttt{SUM=$\sum_{1\leq j \leq q }v_j$}, \qquad v_j \in dom(A) \quad \forall j=1,...,q.
\end{equation*} 

The CSP computes the sum of the
splits stored in $A_i$ as \texttt{SUM}$_i \qquad \forall i=1,\dots k-1$ such that\\

\begin{equation*}%\label{Equation.sumi}
\texttt{SUM}_i= \sum_{1\leq j \leq q, 1\leq i \leq k-1}v_{i,j} \quad \mod p.	
\end{equation*}

Then, \texttt{SUM$_1$}, \texttt{SUM$_2$},..., \texttt{SUM$_{k-1}$} are shipped back to the user and polynomial $P_{\texttt{SUM}}(x)$ is built using Equation~\ref{Equation.Lagrange11} using $k$ points

\begin{equation*}\label{Equation.sum2}
(\mathtt{x_i}, \texttt{SUM$_{i}$})_{ i=1,...,k-1}, \quad (\mathtt{x_k}, \sum_{j=1}^{q}v_k=q \times v_k).
\end{equation*}

$P_{\texttt{SUM}}(x)$'s constant term is \texttt{SUM}.
S4 does not alter the number of records. Hence, \texttt{COUNT} queries can be processed normally, thus also allowing \texttt{AVG} queries.

\subsection{Security Analysis}
\label{Section.S4.Analysis}

Paillier's PHE is semantically secure, but it is too expensive in terms of ciphertext storage space and query response time. S4 proposes a classical trade-off with a lower level of security, but better storage and response time efficiency \cite{shamir1979share}.
Let us consider a scenario where the CSP is said honest but curious, which is a widely used adversary model for cloud data outsourcing \cite{DBLP:conf/sdmw/SanamradK13}. 
Such a CSP faithfully complies to any service-level agreement and, in our particular case, stores data, runs queries and provides results without alteration, malicious or otherwise. Yet, the CSP may access data and infer information from queries and results. 

%The idea behind the use of cryptographic schemes such as Paillier's and S4 is that the CSP should not be fully trusted, since it might be curious or hacked. Paillier's PHE is semantically secure, i.e., ciphertext is indistinguishable from random data. However, it is too expensive in terms of ciphertext storage space and query response time. S4 introduces a new trade-off with a lower level of security, but better storage space and query response time efficiency.

%More precisely, p
Privacy in S4 relies on the fact that a secret value is only retrievable by the user via private key $\mathcal{K}$. As in secret sharing, it is indeed guaranteed that at least $k$ splits and $X$ are necessary to reconstruct a secret, while the CSP has access to only $k-1$ splits. Both $X$ and the $k^{th}$ split, i.e., $\mathcal{K}$, are stored at the user's. However, the CSP still has access to linear combinations of splits (Equation~\ref{Equation.Lagrange11}), 
which provide some information. Still, the higher $k$ is, the more difficult it is to interpret linear combinations of splits. Thus, $k$ is the prime security parameter in S4. Experiments in Section~\ref{Section.Experimental Evaluation} 
provide hints for choosing $k$.

Moreover, if some secrets are known by the CSP, e.g., through public communication of a company to its shareholders, solving Equation~\ref{Equation:LagrangeBasis} 
becomes possible. For example, if the CSP knows secrets $v_1, ..., v_{k-1}$. Also knowing the corresponding splits $v_{1,j}, ..., v_{k-1,j}$ $\forall j \in [1, k-1]$, the CSP can recover the Lagrange basis polynomials $\ell_i(0)$ $\forall i \in [1, k]$ and solve Equation~\ref{Equation:LagrangeBasis} 
to recover all secrets. However, the CSP must know at least $k-1$ secrets to do so. % highlighting again the importance of $k$ for enforcing data security. 
Moreover, we also propose leads to address this problem in Section~\ref{Section.Conclusion}.

\section{Experimental Evaluation}
\label{Section.Experimental Evaluation} 

%In this section, we compare S4 to Paillier's PHE in terms of performance, i.e., encryption and decryption time, query processing time and space overhead.

\subsection{Experimental Setup}  

%\subsubsection{Hardware and Software Configuration}

We implement S4 in C using compiler gcc 4.8.2. S4's source code is freely available on-line\footnote{http://eric.univ-lyon2.fr/download/libS4.zip}. %under free Creative Commons license CC BY-SA \cite{CC-BY-SA4}. 
Experiments related to Paillier's PHE exploit the libpaillier standard C library \cite{libpaillier}. All mathematical computations use the GNU Multiple Precision Arithmetic Library (GMP) \cite{gmp}. 
Eventually, we conduct our experiments on an Intel Core i7 3.10 GHz PC with 16 GB of RAM running Linux Ubuntu 15.05.

%\subsubsection{Datasets} 

We compare S4 and Paillier's PHE using simple synthetic datasets, i.e., %the Employee table from Section~\ref{exs4} with one single Salary attribute. Salaries are 
32-bit unsigned integers generated uniformly at random from the integer range $[10^3, 10^4[$. We scale up the number of records $m$ such that $m \in (10^3$, $10^4$, $10^5$, $10^6$), forming four distinct datasets. %We execute summation queries on each dataset, i.e., the sum of all salaries.

%\subsubsection{Parameter Setup} 

In S4, we vary $k$ from $8$ to $64$, higher values of $k$ inducing too long execution times. Prime $p$ must be greater than the greatest query answer, e.g., $p>\sum_{j=1}^m v_j$.
In Paillier's PHE, we use a key size of 1024 bits, which induces ciphertexts of 2048 bits. Such key size is the absolute minimum to achieve security \cite{DBLP:journals/iacr/JostLMS15,DBLP:journals/jdwm/YildizliPSSL11}. 

\subsection{Encryption and Decryption Time}

%In this first series of experiments, we compare the computation time of S4 for secret splitting/reconstruction against Paillier's encryption/decryption. To this aim, we measure client processing times.
  
\begin{figure}[hbt]
\centering
\begin{minipage}{.5\textwidth}
  \centering
  \includegraphics[width=7cm]{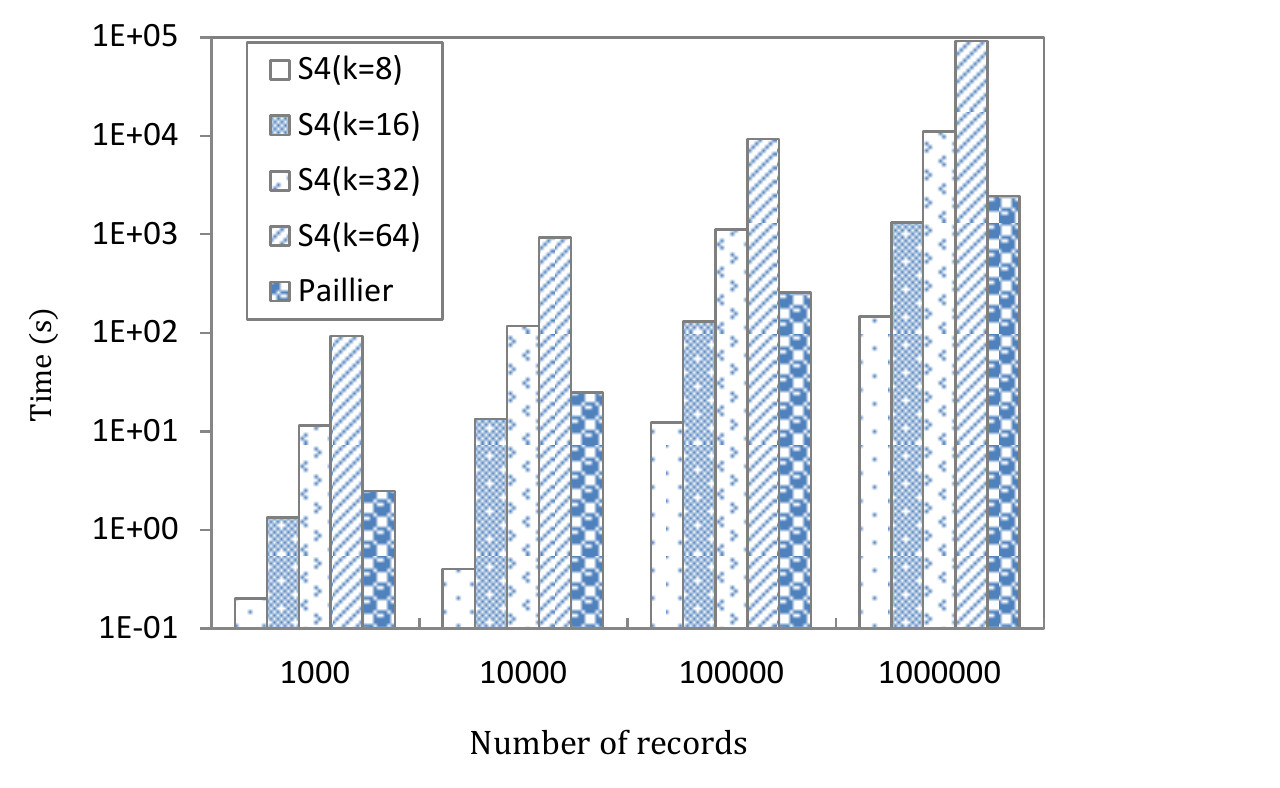}
  \captionof{figure}{Splitting/encryption time}
  \label{fig:Enc}
\end{minipage}%
\begin{minipage}{.5\textwidth}
  \centering
  \includegraphics[width=7cm]{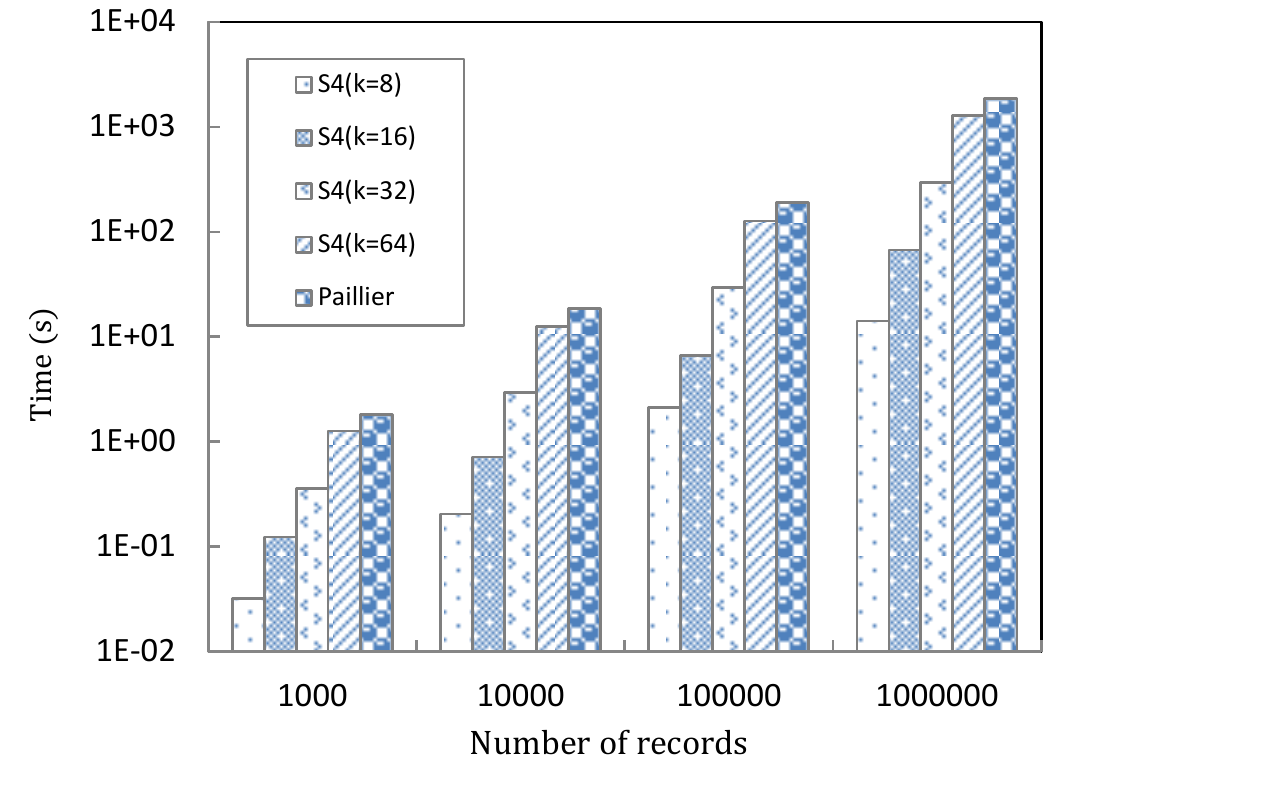}
  \captionof{figure}{Reconstruction/decryption time}
  \label{fig:Dec}
\end{minipage}
\end{figure}

Figure~\ref{fig:Enc} plots the time of secret splitting in S4 and secret encryption in Paillier's scheme with respect to $m$. It shows that encryption time in S4 is lower than Paillier's when $k \leq 16$, and then becomes higher when $k \geq 16$. Secret splitting consists in building a random polynomial % and evaluating it in $k-1$ points. In order to build the random polynomial, 
by randomly choosing $k-2$ points. %must be randomly chosen 
%from $\mathbb{F}_p$. 
Hence, splitting time increases with $k$. Figure~\ref{fig:Enc} actually illustrates the tradeoff between S4's security and encryption efficiency with respect to Paillier's PHE.

\begin{comment}

\begin{center}
	\begin{figure}[hbt]
		%%\vspace{2.85cm}
		\centering
		\includegraphics[width=10.25cm]{New_Enc.pdf}
		\caption{Splitting/encryption time}
		\label{fig:Enc}
	\end{figure}
\end{center} 

\end{comment}
 
Figure~\ref{fig:Dec} plots the time of secret reconstruction in S4 and secret decryption in Paillier's scheme with respect to $m$.	%The situation is different. 
With the selected values of $k$, decryption is faster with S4 than with Paillier's PHE. This is mainly because Paillier's scheme needs $m$ expensive modular multiplications of large, 2048-bit numbers for decryption, while secret reconstruction in S4 works by polynomial interpolation over $k$ points and evaluating the polynomial in one single point. 

\begin{comment}

\begin{center}
	\begin{figure}[hbt]
		%%\vspace{2.85cm}
		\centering
		\includegraphics[width=10.25cm]{New_Dec.pdf}
		\caption{Reconstruction/decryption time}
		\label{fig:Dec}
	\end{figure}
\end{center}

\end{comment}

\vspace{-1cm}
\subsection{Space Overhead}

%In this series of experiments, we compare the sizes of ciphertexts produced by S4 and Paillier's PHE, respectively. 

Figure~\ref{fig:Storage} plots the storage required by S4 and Paillier's PHE with respect to $m$. With the selected values of $k$, S4's  storage overhead is always much smaller than that of Paillier's PHE since Figure~\ref{fig:Storage}'s y axis follows a logarithmic scale. 
Paillier's scheme indeed produces 2048-bit ciphertexts. Thus, its storage overhead is $m \times 2048$.  
With S4, each value is split into $k-1$ values. Thus, S4's storage overhead is $m \times (k-1)$ times plaintext size.

\begin{comment}

\begin{center}
	\begin{figure}[hbt]
		%%\vspace{2.85cm}
		\centering
		\includegraphics[width=9cm]{New_Storage}
		\caption{Storage overhead}
		\label{fig:Storage}
	\end{figure}
\end{center}

\end{comment}

\subsection{Query Processing Time}

%Finally, in this last series of experiments, we compare the efficiency of S4 and Paillier's PHE for query  processing. To this end, we executed summation queries over all records in each dataset. 

\begin{figure}[hbt]
\centering
\begin{minipage}{.5\textwidth}
  \centering
  \includegraphics[width=6.63cm]{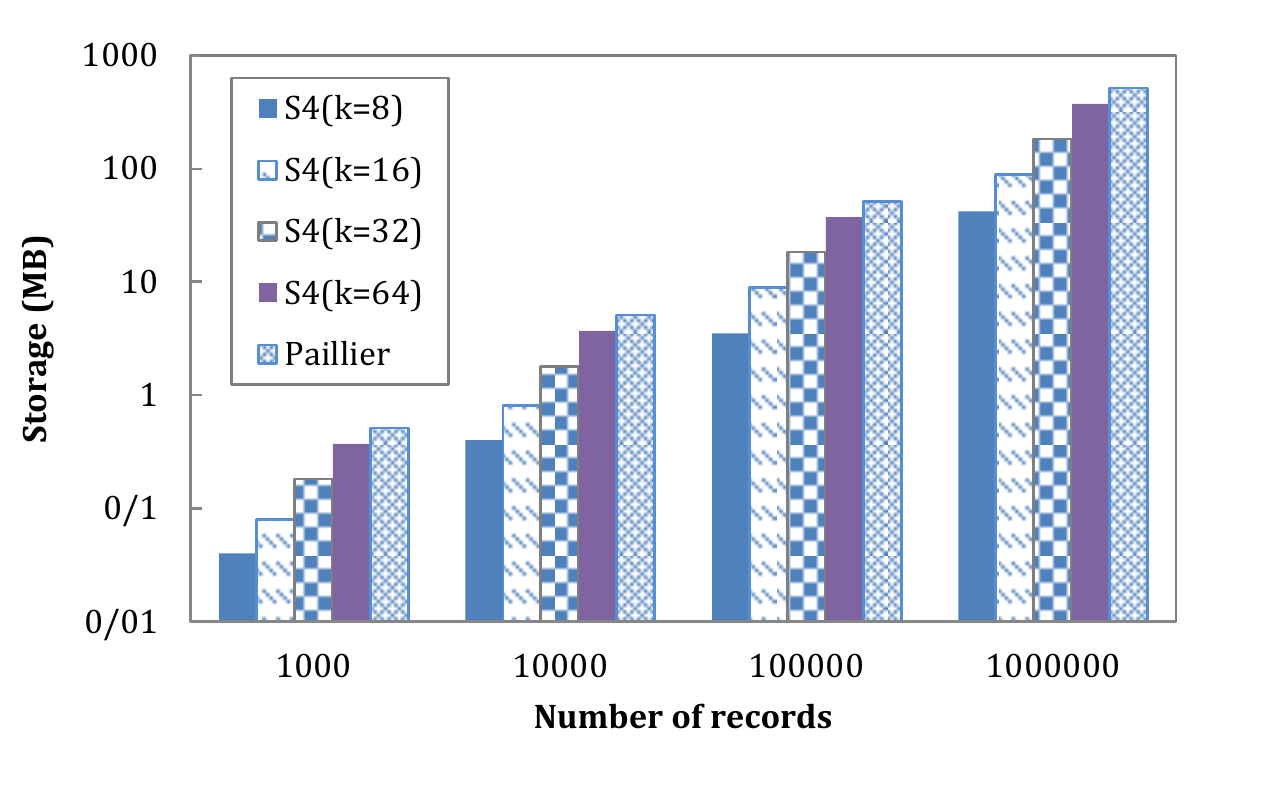}
  \captionof{figure}{Storage overhead}
  \label{fig:Storage}
\end{minipage}%
\begin{minipage}{.5\textwidth}
  \centering
  \includegraphics[width=7cm]{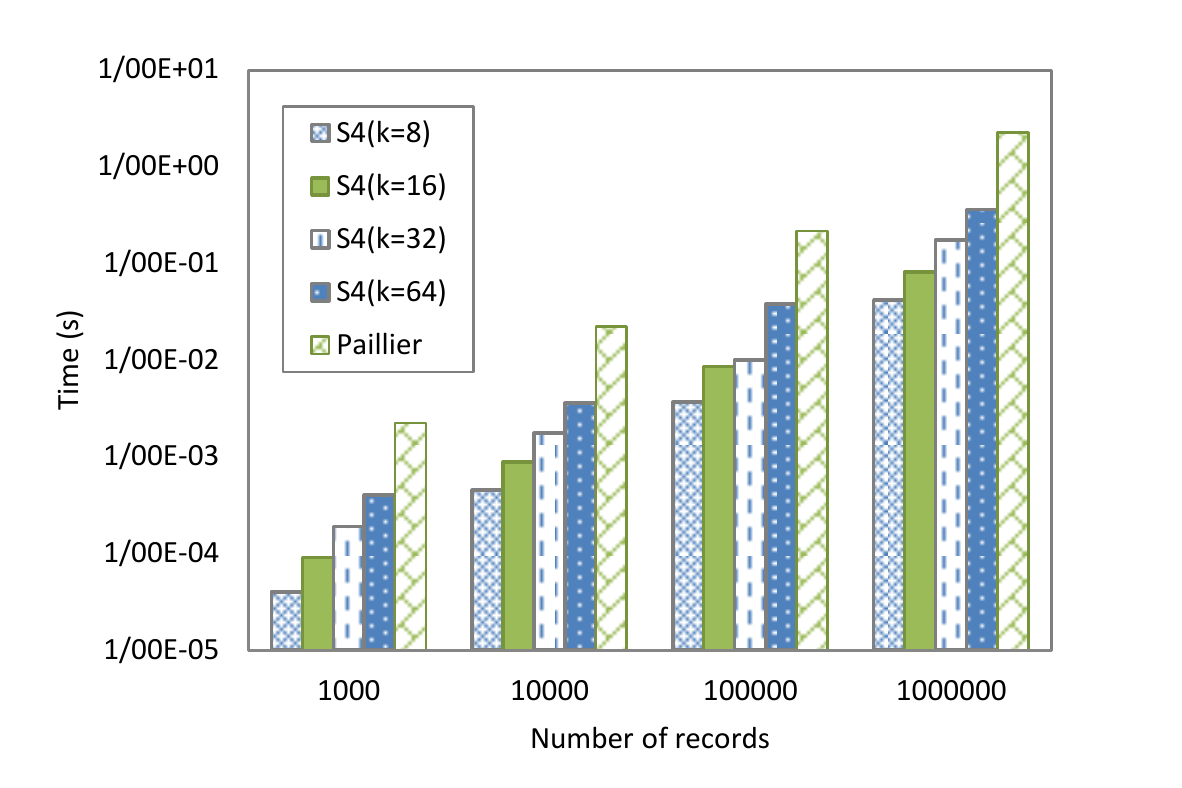}
  \captionof{figure}{Summation execution time}
  \label{fig:Sum}
\end{minipage}
\end{figure}

Figure~\ref{fig:Sum} plots summation query processing times over all records in each dataset, for both S4 and Paillier's PHE, with respect to $m$. It shows that, with the selected values of $k$, query execution time in S4 is lower than that of Paillier's scheme. This is because Paillier's scheme requires $m$ expensive modular multiplications to compute a sum, while S4 computes only $(k-1)\times m$ simple modular additions.  

\begin{comment}

\begin{center}
	\begin{figure}[hbt]
		%%\vspace{2.85cm}
		\centering
		\includegraphics[width=9cm]{New_Sum.pdf}
		\caption{Summation query execution time}
		\label{fig:Sum}
	\end{figure}
\end{center}

\end{comment}

\vspace{-1cm}
\section{Conclusion}
\label{Section.Conclusion}

In this paper, we introduce S4, a new cryptographic scheme that supports summation queries in cloud-based OLAP. We experimentally show that S4 is much more efficient than Paillier's PHE in terms of query response time and space overhead. Thus, replacing Paillier's scheme with S4 in secure DBMSs such as CryptDB and MONOMI can improve analytical query processing in cloud DWs. Moreover, we also plan a variant of S4 for computing multiplications. % over encrypted data.

However, we achieve performance gains through a slight degradation of security, especially when an adversary has knowledge of secret values. Although it is definitely acceptable in some cloud DW and OLAP scenarios, e.g., public aggregate data might not actually yield secrets, i.e., fine-grained data, we will devote future research to strengthen S4 against such threats.
More precisely, we plan to introduce noise, as in many cryptographic problems such as %approximate Greater Common Divisor (
approximate-GCD \cite{DBLP:journals/iacr/Liu15} or %Learning With Errors (
LWE \cite{DBLP:journals/jacm/Regev09}. For instance, instead of sharing $v_j$, we could share $10^r\times v_j+noise$. By doing so, security is intuitively enhanced  while the whole process remains correct, provided $r$ is sufficiently large and $noise$ sufficiently small.

\bibliographystyle{splncs03}
\bibliography{cloud_crypt}

\end{document}